\documentclass[aps,prl,groupedaddress,draft]{revtex4}
\begin{document}


\voffset1.5cm
\title{More on eikonal approximation for high energy scattering.}
\author{Tolga Altinoluk, Alex Kovner and Javier Peressutti}
\affiliation{ Physics Department, University of Connecticut, 2152 Hillside road, Storrs, CT 06269, USA}

\date{\today}
\begin{abstract}
We formulate eikonal approximation to the calculation of high energy scattering amplitude in the frame where both colliding objects are very energetic. We express the eikonal scattering matrix in terms of the color charge densities of the colliding objects. The calculation is performed in the Hamiltonian formalism. We also show that the appearance of the longitudinal electric and magnetic fields immediately following the collision is fully taken into account in the eikonal approximation.
\end{abstract}
\maketitle

\section{Introduction.}
Partonic level eikonal approximation has been extensively used in recent years in the study of high energy evolution of hadronic scattering \cite{JIMWLK}. As explained in \cite{one}, this approximation assumes  only integrity of partons throughout the scattering event rather than the integrity of the whole hadronic state. It thus allows for inelastic rather than only elastic final states. 
The approximation is most convenient to use  in the situation where one of the colliding hadrons carries almost all the energy of the collision. In this case the picture of the scattering is that of the partons from the wave function of the fast hadron propagating through the color fields associated with the target - an almost static hadron\cite{zakopane}.
This setup is very convenient when the scattering is strongly asymmetric, namely when the "projectile" hadron is small and contains a small number of partons, while the target hadron is large and is characterized by strong fields. Hamiltonian description in this case is given straightforwardly in terms of the light cone wave function of the projectile \cite{zakopane},\cite{klwmij}.

On the other hand in a more symmetric situation, such as scattering of two small objects or two large objects, like nuclei, it is advantageous to describe the scattering event in a symmetric frame where both colliding objects have comparable energies. This requires the use of a symmetric gauge condition. In Lagrangian formulation this has been done for both "p-p" and "A-A" scattering \cite{balitsky}, \cite{mkw},\cite{im}. We are however not aware of a Hamiltonian formulation of this problem in a symmetric frame and gauge. It is desirable to have such a formulation, since Hamiltonian approach proved to be very fruitful and intuitive in the study of high energy scattering\cite{klwmij}, \cite{duality},\cite{remarks},\cite{foam}. 

Our purpose in the first part of this paper is to provide such a Hamiltonian formulation. We use a symmetric gauge $A_0=0$ with the subsidiary gauge fixing condition $A_3(t=0)=0$ and show how to calculate the eikonal $S$-matrix in the "classical approximation", namely in the situation when the filds of both incoming hadrons are parametrically large. 

The second part of this paper is a short comment on the nature of the longitudinal fields that are created in the initial stages after the collision. The fact of the appearance of such fields has been realized only in recent years and this realization prompted a lot of discussion\cite{long}. It has been proposed for example that they may be important in the dynamics of the formation of so called "ridge" in heavy ion collisions\cite{glasma}.
Given that the eikonal approximation has been extensively used in the study of hadronic scattering at high energies, it is natural to ask whether these longitudinal fields are accounted for in the eikonal approximation, or whether they are purely non eikonal effect. If the latter were the case, it would cast serious doubt on the applicability of the eikonal approximation to the nucleus-nucleus collision. It seems to be an implicit and unspoken assumption of the community at present, that longitudinal fields are indeed outside the eikonal approximation, even though this question has not been studied explicitly. We point out in this paper that the actual situation is in fact the opposite. By analyzing the structure of the hadronic wave function immediately after (eikonal) scattering we show  that the longitudinal fields arise naturally and unavoidably in the "wake" of the projectile. The reason for this is that the Weiszacker-Williams gluons of the projectile do not gauge rotate "in sinc" with the valence color charges in the projectile wave function. This disturbs the balance necessary to preserve the absence of the longitudinal fields. Thus eikonal scattering takes properly into account this important piece of dynamics.

\section{Eikonal Approximation in the Symmetric Gauge.}
We are aiming to study the scattering of two hadrons, where one of them (B) is moving right and the other one (C) moving left with high longitudinal momentum. Most of the partons in the wave functions of those two objects therefore have very large positive or negative longitudinal momenta.
To discuss the scattering in a symmetric manner we need first off to choose a symmetric gauge. The kinematics of the process suggests that we choose
\begin{equation}
A_0=0
\end{equation}
In this gauge the theory is defined by the Hamiltonian density
\begin{equation}
  \mathcal{H} = \frac{1}{2}(\Pi^i)^2 + \frac{1}{4}(F_{ij})^2
\end{equation}
with the constraint
\begin{equation}
  D\cdot\Pi = 0
\end{equation}
where $\Pi^a_i(x)$ are momenta (color electric field) conjugate to the vector potential $A^a_i(x)$, the color magnetic field $F^{a}_{ij}(x)=\partial_iA^a_j-\partial_jA^a_i-gf^{abc}A^b_iA^c_j$, and $D^{ab}_i=\partial_i\delta^{ab}-gf^{acb}A^c_i$.

We also impose a subsidiary gauge fixing condition to fix the gauge transformations that do not depend on time
\begin{equation}\label{subs}
A_3({\mathbf x},t=0)=0
\end{equation}
Imposition of the subsidiary condition in the Hamiltonian formalism is not frequently discussed. We have provided Appendices A and B to explain how this is done conceptually and also technically. The bottom line is that for our purposes, as long as we work in the classical approximation where the fields associated with the incoming hadrons are large, the subtleties of the procedure are not relevant, and we can treat eq.(\ref{subs}) as operatorial constraint which can be used to eliminate $A_3({\mathbf x})$.

Fixing $A_3(t=0)=0$, we can solve for $\Pi^3$ from the Gauss' law:
\begin{equation}
  \Pi^3 = -\frac{1}{\partial_3} D\cdot \Pi
\end{equation}
and the Hamiltonian becomes
\begin{equation}
  \mathcal{H} = \frac{1}{2}(\Pi^i)^2 + \frac{1}{2}(\Pi^3)^2 + \frac{1}{4}(F_{ij})^2  +\frac{1}{2}(F_{i3})^2 \label{ham}
\end{equation}
where from now on we take the index $i$ to take two values $i=1,2$.
Our residual gauge fixing is not complete as it does not fix gauge transformations which do not depend on $x_3$. Thus the theory still has a constraint
\begin{equation}\label{resgaus}
\int dx_3D_i\Pi_i=0
\end{equation} 

The fields and momenta satisfy the canonical commutation relations
\begin{equation}
  [A_i^a(\mathbf{x}),\Pi_b^j(\mathbf{y})]=i\delta_{ib}^{ja}(\mathbf{x}-\mathbf{y})    \label{crAPI}
\end{equation}
and can be expressed in terms of the canonical creation and annihilation operators $a_i$ and $a_i^{\dagger}$ as
\begin{eqnarray}
   A^a_i(\mathbf{x}) &=& \int \frac{d^3k}{(2\pi)^3}  \frac{1}{\sqrt{2\omega_{\mathbf{k}}}} \, \big[  a^a_i(\mathbf{k}) e^{i\mathbf{k}\cdot\mathbf{x}}  +
  																																											a^{\dagger a}_i(\mathbf{k}) e^{-i\mathbf{k}\cdot\mathbf{x}}   \big]   \label{A}\\
  \Pi^a_i(\mathbf{x}) &=& -i\int \frac{d^3k}{(2\pi)^3} \sqrt{\frac{\omega_{\mathbf{k}}}{2}} \, \big[  a^a_i(\mathbf{k}) e^{i\mathbf{k}\cdot\mathbf{x}}  +
  																																											a^{\dagger a}_i(\mathbf{k}) e^{-i\mathbf{k}\cdot\mathbf{x}}   \big]   \label{PI}
\end{eqnarray}  																																											
The canonical commutation relations are as always
\begin{equation}
  [a(\mathbf{k}),a^{\dagger}(\mathbf{p})] = (2\pi)^3 \delta^{ab}_{ij}\delta^3(\mathbf{k}-\mathbf{p})    \label{craad}
\end{equation}
For simplicity of notation we will omit the color and rotational indices on the fields in the following.

We now artificially split the vector potential $A$ into the sum of three fields in a manner similar to \cite{balitsky}
\begin{equation}\label{split}
A=B+C+\widetilde{A}
\end{equation}
 where 
\begin{eqnarray}
  B(\mathbf{x}) &=& \int_{\Lambda}^{\infty} \frac{d^3k}{2\pi} \frac{1}{\sqrt{2\omega_k}} \, \big[  a(k) e^{ikx} + a^{\dagger}(k) e^{-ikx}   \big]   \label{B}\\
  C(\mathbf{x}) &=& \int_{-\infty}^{-\Lambda} \frac{d^3k}{2\pi} \frac{1}{\sqrt{2\omega_k}} \, \big[  a(k) e^{ikx} + a^{\dagger}(k) e^{-ikx}   \big]   \label{C}\\
  \widetilde{A}(\mathbf{x}) &=& \int_{-\Lambda}^{\Lambda} \frac{d^3k}{(2\pi)^3}  \frac{1}{\sqrt{2\omega_{\mathbf{k}}}} \, 
             \big[  a(\mathbf{k})  e^{i\mathbf{k}\cdot\mathbf{x}}  +	a^{\dagger}(\mathbf{k}) e^{-i\mathbf{k}\cdot\mathbf{x}}   \big]   \label{At}
\end{eqnarray}
where the limits in the above integrals refer to the variable $k_3$. 
The reasoning behind this splitting is the following. We do not intend to discuss all states in the Hilbert space of the QCD Hamiltonian, but only those states which contain fast moving hadrons. The wave function of such a hadron contains overwhelmingly gluons with very large (positive or negative) momentum $k_3$. Thus the splitting in eq.(\ref{split}) corresponds to separating the degrees of freedom describing the right moving hadron (fields $B$) and the left moving hadron (fiields $C$). The field $\tilde A$ contains the soft gluons which necessarily appear in the final state of the scattering and thus generally have to be kept. Technically they are also necessary to satisfy the Gauss' law eq.(\ref{resgaus}) 

The longitudinal cutoff $\Lambda$ is arbitrary and its only role is to indicate that the longitudinal momenta in the fields $B$, $C$ and $\tilde A$ are vastly different. We will be thinking of it as being of the order of typical transverse momentum $k_\perp$. Since most of the momenta in $B$ and $C$ are much larger than $\Lambda$, we can approximate the energy $\omega_k$  entering eqs.(\ref{B},\ref{C},\ref{At}) by $\omega_k\approx |k_3|$. We thus have
\begin{eqnarray}
  B(\mathbf{x}) &=& \int_{\Lambda}^{\infty} \frac{d^3k}{2\pi} \frac{1}{\sqrt{2|k_3|}} \, \big[  a(k) e^{ikx} + a^{\dagger}(k) e^{-ikx}   \big]   \label{B1}\\
  C(\mathbf{x}) &=& \int_{-\infty}^{-\Lambda} \frac{d^3k}{2\pi} \frac{1}{\sqrt{2|k_3|}} \, \big[  a(k) e^{ikx} + a^{\dagger}(k) e^{-ikx}   \big]   \label{C1}\\
  \widetilde{A}(\mathbf{x}) &=& \int_{-\Lambda}^{\Lambda} \frac{d^3k}{(2\pi)^3}  \frac{1}{\sqrt{2\omega_{\mathbf{k}}}} \, 
             \big[  a(\mathbf{k})  e^{i\mathbf{k}\cdot\mathbf{x}}  +	a^{\dagger}(\mathbf{k}) e^{-i\mathbf{k}\cdot\mathbf{x}}   \big]   \label{At1}
\end{eqnarray}
Unsurprisingly, the expressions for $B$ and $C$ now become exactly the same as in the appropriate light cone gauges - the $A^+=0$ for the field $B$ and the $A^-=0$ for the field $C$. It is not particularly surprising, since in our gauge both conditions hold at time zero.
Splitting the conjugate momentum in the similar way we have
\begin{equation}
\Pi = \Pi_B + \Pi_C + \Pi_{\widetilde{A}}
\end{equation}
with
\begin{eqnarray}
  \Pi_B(\mathbf{x}) &=& -i\int_{\Lambda}^{\infty} \frac{d^3k}{2\pi} \sqrt{\frac{|k_3|}{2}} \, \big[  a(k) e^{ikx} - a^{\dagger}(k) e^{-ikx}   \big]   \label{PIB}\\
  \Pi_C(\mathbf{x}) &=& -i\int_{-\infty}^{-\Lambda} \frac{d^3k}{2\pi} \sqrt{\frac{|k_3|}{2}} \, \big[  a(k) e^{ikx} - a^{\dagger}(k) e^{-ikx}   \big]   \label{PIC}\\
  \Pi_{\widetilde{A}}(\mathbf{x}) &=& -i\int_{-\Lambda}^{\Lambda} \frac{d^3k}{(2\pi)^3}  \sqrt{\frac{\omega_{\mathbf{k}}}{2}} \, 
             \big[  a(\mathbf{k})  e^{i\mathbf{k}\cdot\mathbf{x}}  -	a^{\dagger}(\mathbf{k}) e^{-i\mathbf{k}\cdot\mathbf{x}}   \big]   \label{PIAt}
\end{eqnarray}
The above equations give
\begin{equation}
  \Pi = -\partial_3 B + \partial_3 C + \Pi_{\widetilde{A}}
\end{equation}

Our next step is to express the Hamiltonian in terms of the fields $B,C$ and $\tilde A$ again in the approximation where the longitudinal momenta in the fields $B$ and $C$ are much greater than the transverse momenta. Also we are interested mainly in the case when the intensity of the fields $B$ and $C$ is large. More specifically we are gearing up to the "classical" situation where the number of gluons in each of the incoming hadrons is of order $1/ g^2$.
\begin{equation}
\int_{-\Lambda}^\infty dk_3d^2k_\perp a^\dagger(k)a(k)\sim \int_{-\infty}^{\Lambda} dk_3d^2k_\perp a^\dagger(k)a(k)\sim {1\over g^2}
 \end{equation}
 This defines our parametric counting. Assuming that the integral over longitudinal momenta is saturated on some large momentum scale $\kappa$, we have
 \begin{equation}
 a(k)\sim a^\dagger(k)\sim {1\over \sqrt {\kappa}}{1\over gk_\perp}
 \end{equation}
 Using this counting and keeping only the leading (order $\kappa$) and the first subleading terms (order $\kappa^0$) in the Hamiltonian we find
 \begin{equation}
 H=H_{0}+H_1
 \end{equation}
 with
 \begin{equation}
 H_{0}=\int d^3x \left[\left(\partial_3B\right)^2+\left(\partial_3C\right)^2\right]
 \end{equation}
 and
 \begin{equation}\label{h1}
 H_1=\frac{1}{2}(\Pi^i)^2 +\frac{1}{4}(F_{ij})^2  +\frac{1}{2}(\partial_3A_i)^2+{1\over 2}\left[{1\over \partial_3}\left(D_i\Pi_i+gf^{abc}\partial_3B_i^bB_i^c +g f^{abc}\partial_3C_i^bC_i^c\right)\right]^2 
 \end{equation}
 
 where we have dropped the tilde over $A$ and $\Pi$; and $D$ denotes covariant derivative in the soft field $A$.
 
 The residual Gauss' law eq.(\ref{resgaus}) becomes
 \begin{equation}
 \int dx_3D_i\Pi_i+J^++J^-=0
 \end{equation}
 with
 \begin{equation}
 J^+_a(x_i)=g\int dx_3 f^{abc}\partial_3B_i^bB_i^c; \ \ \ \ \ \ \ \  J^-_a=g\int dx_3 f^{abc}\partial_3C_i^bC_i^c
 \end{equation}
 
 Our goal now is to calculate the eikonal $S$-matrix for the scattering process of two large hadrons (nuclei) in terms of the light cone currents $J^+$ and $J^-$. As the first step towards this goal we will rewrite the Hamiltonian in the interaction picture.
\section{The interaction picture Hamiltonian.}
The interaction picture of quantum mechanics is defined with respect to some free Hamiltonian. In our case it is natural to choose as a free Hamiltonian $H_0$ as it is the leading operator in the high energy limit.

Any operator $O$ in the interaction picture is defined as :
\begin{equation}   O(t)=U_0^{\dagger}(t,t_0)O(t_0)U_0(t,t_0) 
\end{equation}
where $U_0$ is the evolution operator, and $H_0$ is the free Hamiltonian:
\begin{equation} U_0(t,t_0)=e^{-iH_0(t-t_0)},  \quad  H_0=\int_x \big[(\partial B)^2+(\partial C)^2\big]   
\end{equation}
The free Hamiltonian $H_0$ is just the Hamiltonian of free left and right moving energetic partons
\begin{eqnarray}
H_0&=&H_B+H_C\nonumber\\
  H_B &=& \int_{k_3>0}  \frac{d^3k}{2\pi}\, |k_3|\, a^{\dagger}(k)\cdot a(k)   \label{HB}\\
  H_C &=& \int_{k_3<0}  \frac{d^3k}{2\pi}\, |k_3|\, a^{\dagger}(k)\cdot a(k)   \label{HC}
\end{eqnarray}
The interaction picture creation and annihilation operators are
\begin{eqnarray*}
  a(k,t) &=& a(k)\,e^{-i\,|k_3|\,t}  \\
  a^{\dagger}(k,t) &=& a^{\dagger}(k)\,e^{i\,|k_3|\,t}
\end{eqnarray*}


Since $H_0$ does not depend on field $A$, the interaction picture and the Schroedinger picture $A$ and $\pi$ are identical. To calculate the interaction picture Hamiltonian $H_1$ we need to look at the terms of the type 
\begin{equation}
  B(-k-p,t)\cdot B(p,t)
  \end{equation}
 with momentum $k_3\sim k_\perp$. For the Schroedinger operators we could neglect $k_3$ in this expression, $k_3+p_3\approx p_3$, since expanding in $k_3$ necessarily meant expanding in $k_3/p_3$. Now however there is an additional dimensional parameter in the game, time $t$. We thus have to be careful not to neglect terms of the type $k_3t$ which do not have to be small.

Keeping this in mind we write
\begin{equation}
  B(-k-p,t)\cdot B(p,t) =
   \frac{1}{2|p_3|}\,   \Big\{     \theta(-p_3)\,a(-k-p)\cdot a^{\dagger}(-p)\,  e^{-i\{|k_3+p_3|\, - |p_3|\}t}  
+      \theta(p_3)\,a^{\dagger}(k+p)\cdot a(p)\,  e^{i\{|k_3+p_3|\, - |p_3|\,\}t}   \Big\}
\end{equation}
For $k_3\ll p_3$,
\begin{equation}  |k_3+p_3|\,-|p_3|\,  \simeq  k_3\,{\rm sign}(p_3)   \end{equation}
Referring to eq.(\ref{h1}) we see that we need to calculate
\begin{equation}
{1\over\partial_3}gf^{abc}\partial_3B_i^bB_i^c
\end{equation}
or in momentum space
\begin{equation}
\int dp_3{p_3\over k_3}gf^{abc}B_i^b(-k_3-p_3)B_i^c(p_3)=g{1\over k_3}e^{ik_3t}f^{abc}\int_0^\infty dp^3a^{\dagger b}_i(p)a^c_i(p)
\end{equation}
Analogous expression holds for the term involving the fields $C$ with the substitution $k_3\rightarrow-k_3$.
All said and done we obtain the "interaction" Hamiltonian in the interaction picture as

\begin{equation}
  H_I = \frac{1}{2} \int_x  \bigg\{     \Big[ \frac{1}{\partial_3}\big( D_i\cdot\Pi_i + \delta(x_3-t) J^+ + \delta(x_3+t) J^- \big)\Big]^2
  							+\Pi_i^2 + \frac{1}{2}F_{ij}^2 + (\partial_3 A_i)^2              \bigg\}
\end{equation}
This expression is indeed what one naively expects. Since the most energetic "valence" partons are very fast, in the leading eikonal approximation their Hamiltonian is free. On the other hand the soft modes of the field  $A$ in the leading order are eikonally coupled to the right- and left- moving thin pancakes of color charge associated with the fast modes $B$ and $C$.

Our aim now is to calculate the S-matrix in the parametric regime $J^+\sim J^-\sim {1\over g}$. 
In this classical regime the $S$-matrix is given by the semiclassical expression
\begin{equation}\label{s}
S=\exp\left\{i\int_{-\epsilon}^\epsilon dt  \int \!d^3x \big( \dot{A}\cdot\Pi - H \big)  \right\}
\end{equation}
The time integral in this expression is only over the interaction time, which in the eikonal approximation is infinitesimal. Thus to calculate the $S$-matrix we need to find the classical solution of equations of motion only up to infinitesimaly short time after the interaction. 
The equations of motion that follow from the Hamiltonian are
\begin{eqnarray}
  \dot{A}^a_i &=& \Big[  D_i\frac{1}{\partial_3^2}  \big( D\cdot\Pi + \delta(x_3-t) J^+ + \delta(x_3+t) J^- \big) \Big]^a    +   \Pi^a_i   \label{eqsmotion} \\
  \dot{\Pi}^a_i &=& -g f^{abc} \Pi^b_i \frac{1}{\partial_3^2}  \big[ D\cdot\Pi + \delta(x_3-t) J^+ + \delta(x_3+t) J^- \big]^c
  											+  \big( D_j F_{ji}\big)^a  +  \partial_3^2 A^a_i          \nonumber
\end{eqnarray}
For $t<0$ the solution for given $J^+$ and $J^-$ is
\begin{eqnarray}\label{sol-}
  A^i_{0 a}&=& -\theta(-x_3+t)b^{+i}_a - \theta(x_3+t)b^{-i}_a  \\
  \Pi^i_{0 a} &=& -\delta(-x_3+t)b^{+i}_a - \delta(x_3+t)b^{-i}_a
\end{eqnarray}
with the "classical" fields $b_i^\pm $ defined by
\begin{eqnarray}
&& \partial_i b^+_i = J^+   \ ; \ \ \ \ \ \   \partial_i b^-_i = J^- \nonumber\\
&&F^{ij}(b^+)=F^{ij}(b^-)=0
\end{eqnarray}  
Since the time integral in eq.(\ref{s}) is over the infitesimal interval, only discontinuities of the type $\theta(t)$ in $A$ and singularities of the type $\delta(t)$ in $\Pi$ can give finite contributions. It is easy to see that the only discontinuities arise from the extension of eq.(\ref{sol-}) to positive times.
We write the solution for arbitrary time as
\begin{eqnarray}
  A(\mathbf{x},t^a_i) &=& A^a_{0 i}(\mathbf{x},t) + \delta A^a_i\, \theta(t)  \\
  \Pi^a_i(\mathbf{x},t) &=& \Pi^a_{0 i}(\mathbf{x},t) + \delta \Pi^a_i\, \theta(t)
\end{eqnarray}
Substituting this into the equations of motion (\ref{eqsmotion}), we obtain
\begin{eqnarray}\label{api}
{\delta  \dot A}^a_i(\mathbf{x},t)& =&  \delta \Pi^a_i  +  D^{ab}_i(A_0)\frac{1}{\partial_3^2}  \Big[ D^{bc}(A_0)\cdot\delta\Pi^c -gf^{bcd}\delta A^c\Pi_0^d\Big]\\ \label{api1}
 {\delta \dot\Pi}^a_i (\mathbf{x},t)&=& -g f^{abc} \Pi^b_{0i} \frac{1}{\partial_3^2}  \Big[ D^{cd}(A_0)\cdot\delta\Pi^d -gf^{cde}\delta A^d\Pi_0^e\Big]+ \partial_3^2 \delta A^a_i 						+  \Big[ D(A_0)\delta^{ij}-D_i(A_0)D_j (A_0)\Big]^{ab}\delta A^b_j
\end{eqnarray}
These equations have to be solved with vanishing initial conditions
\begin{equation}
\delta A(t=0)=\delta \Pi(t=0)=0
\end{equation}
Examining the right hand side of eqs.(\ref{api},\ref{api1}) we see that they do not have singular terms. The structure of these equations is consistent with the solutions being  regular functions. We thus conclude that $\delta A$ has no $\theta(x_3\pm t)$ type terms and $\delta \Pi$ has no $\delta(x\pm t)$ type terms. 

The classical action therefore can be calculated simply using eq.(\ref{sol-}). The resulting $S$-matrix is
\begin{equation}\label{smatrix}
  S = e^{i \int \!d^2 x\, b_i^+\!\cdot b_i^- }
\end{equation}
This is the same expression as derived in \cite{balitsky} in the path integral formalism.
Note that if one of the color charge densities is small, this expression can be simplified. Let's take for example $J^+\sim g$. Then to leading order in $g$ we have $b^+_i={\partial_i\over \partial^2}J^+$. Substituting this into eq.(\ref{smatrix}) and integrating by parts we find
\begin{equation}
S=e^{i \int \!d^2xd^2y J^+(x){1\over \partial^2}(x,y) J^-( y) }
\end{equation}
which is the expression used in \cite{im} for dipole-dipole scattering.

\section{The eikonal approximation and the longitudinal fields}
The point made in several papers recently\cite{long}, is that the solutions of classical equations after collision contain longitudinal electric and magnetic fields. This is significantly different from the fields incoming into the collision as those are purely transverse. This feature is of course reproduced by our analysis of the previous section. A short time after the collision the solution (up to small corrections) is given be eq.(\ref{sol-}). Calculating the longitudinal electric field we find
\begin{eqnarray}\label{long}
E^a_3&=&\frac{1}{\partial_3} \left[D\cdot \Pi+\delta(x_3-t) J^+ + \delta(x_3+t) J^-\right]=-
gf^{abc}b^{+b}_i(x)b^{-c}_i(x)\theta(x_3+t)\theta(-x_3+t)\theta(t)\nonumber\\
B^a_3&=&{1\over 2}\epsilon_{ij}F^a_{ij}=- gf^{abc}\epsilon_{ij}b^{+b}_i(x)b^{-c}_j(x)\theta(x_3+t)\theta(-x_3+t)\theta(t)
\end{eqnarray}
Thus, immediately  after the collision, the space between the receding fast particles is filled with the longitudinal electric and magnetic fields. 

Since the realization that the longitudinal fields arise after the collision, it has not been entirely clear whether eikonal approximation takes these effects into account. Our discussion in the previous section illustrates that this is not the case, as the longitudinal fields do indeed arise in the eikonal approximation. Conventionally however the  eikonal approximation is formulated in a different frame. In this frame only one of the scattering objects (" the projectile") is fast moving, while the other one is represented by a distribution of static color fields which provide the target for the scattering of the partons from the projectile. It is instructive to understand how the eikonal evolution in this target rest frame also leads to existence of longitudinal fields shortly after collision.

The dynamics of the collision in the target rest frame is the following. The right moving projectile starts up with the wave function of the form:
\begin{equation}
\vert \Psi\rangle_{in}=\Omega[a,a^\dagger,J^+]\vert\psi\rangle_{J^+}
\end{equation}
Here $\vert\psi\rangle_{J^+}$ is the wave function which depends only on the valence charge density, while the unitary operator $\Omega$ depends on the soft gluon degrees of freedom $a$ and $a^\dagger$ and also on $J^+$ as a parameter. The soft gluon degrees of freedom in this approach play the role analogous to that of the soft field $A$ of the previous section, while the charge density $J^+$ is due to the fast modes, the analogs of the field $B$. There are no dynamical modes $C$, as they are represented by the distribution of static color fields $\alpha$. The  operator $\Omega$ is a Bogoliubov type operator as discussed in \cite{foam}. In order to simplify our discussion however, we will consider a classical approximation where it reduces to a coherent operator
\begin{equation}\label{ome}
\Omega=\exp\left\{i\int d\eta d^2x b^a_i(x)\Big[a^a_i(\eta,x)+a^{\dagger a}_i(\eta,x)\Big]\right\}
\end{equation}
where the "classical field $b$ as before, is related to the color charge density via
\begin{equation}\label{b}
\partial_i b^a_i(x) =J^+(x)
\end{equation}
and crucially is two dimensionally a pure gauge
\begin{equation}
\partial_ib^a_j-\partial_jb^a_i-gf^{abd}b^b_i(x)b^c_j(x)=0
\end{equation}
The operators $a^\dagger(\eta)$ and $a(\eta)$ are creation and annihilation operators of soft gluons at rapidity $\eta$.

The operator $\Omega$ when acting on the soft gluon vacuum creates the soft fields of the form
\begin{equation}
A^a_i=-b^a_i(x)\theta(-x^-)
\end{equation}
The transverse electromagnetic field is:
\begin{equation}\label{trf}
F^{+i}_a=b^i_a(x)\delta(x^-)
\end{equation}
The longitudinal electric field in the light cone quantization is not an independent degree of freedom, but is expressed in terms of the transverse components of the vector potential as
\begin{equation}\label{longf}
E_3={1\over \partial^+} \left[D_iF^{+i}-J^+\delta(x^-)\right]
\end{equation}
This solves the Gauss' law in the light cone gauge.
With the transverse field of eq.(\ref{trf}), the eq.(\ref{b}) ensures that the longitudinal electric field before the collision vanishes. The longitudinal magnetic field obviously vanishes as well since $b_i$ is two dimensionally a pure gauge. Thus the incoming projectile contains only transverse electromagnetic fields which are concentrated within a "shock wave" - the transverse plain moving along with the valence charge density created by the fast partons. Those are of course the well known Weizsacker - Williams fields.

The target in this approach is represented by an ensemble of color fields $\alpha^a$. As discussed at length in \cite{klwmij},\cite{remarks},\cite{zakopane} the projectile emerges from the interaction region with the wave function
\begin{equation}
\vert \Psi\rangle_{out}=\Omega[Sa,Sa^\dagger,SJ^+]\vert\psi\rangle_{SJ^+}
\end{equation}
where the single gluon $S$-matrix $S$ is expressed in terms of the target field
\begin{equation}
S(x)=\exp\{i\int dx^-T^a\alpha^a(x,x^-)\}
\end{equation}
For the operator $\Omega$ of eq.(\ref{ome}) this gives
\begin{equation}\label{ome1}
\Omega[Sa,Sa^\dagger,SJ^+]=\exp\left\{i\int d\eta d^2x \bar b^a_i(x)\Big[a^a_i(\eta,x)+a^{\dagger a}_i(\eta,x)\Big]\right\}
\end{equation}
with
\begin{equation}
\bar b_i^a=S^\dagger b^a_i[SJ^+]
\end{equation}
The transformed operator $\Omega$ now creates the transverse field
\begin{equation}
F^{+i}_a=\bar b^i_a(x)\delta(x^-)
\end{equation}
This at first sight looks very similar to the fields before the collision. There is however a crucial difference. These transverse fields accompany the rotated color charge density $SJ^+$. Obviously the color charge density is not rotated entirely "in sinc" with the field in the sense that 
\begin{equation}
\partial b_i\ne SJ^+
\end{equation}
Therefore Gauss' law requires nonvanishing of the longitudinal electric field. 
The longitudinal electric field after the scattering is indeed given by
\begin{equation}\label{eproj}
E_3={1\over \partial^+}\left[D_iF^{+i}-SJ^+\delta(x^-)\right]=\theta(t-x_3)\partial_i\Big(S^\dagger b_i[SJ^+]-b_i[SJ^+]\Big)
\end{equation}
and is clearly nonzero.
Likewise, since $\bar b_i$ is not a two dimensional pure gauge, the longitudinal magnetic field does not vanish either:
\begin{equation}\label{bproj}
F^a_{ij}=\Big[\partial_i\bar b^a_j-\partial_j\bar b^a_i-gf^{abd}\bar b^b_i(x)\bar b^c_j(x)\Big]\theta(t-x_3)
\end{equation}
One note of caution is that eqs.(\ref{eproj},\ref{bproj}) are valid only close to the longitudinal position of the projectile - that is only between the projectile and the target.  When extended beyond the longitudinal coordinate of the target, the integral $1/\partial^+$ will pick up additional contributions due to the target fields, so that presumably the longitudinal fields vanish for negative $x_3$.
Otherwise eqs.(\ref{eproj},\ref{bproj}) closely parallel the structure of longitudinal fields in the symmetric frame.

\section {Appendix A. 
Gauge fixing in the Hamiltonian formalism.}
In this Appendix we discuss how to consistently impose a gauge fixing for the residual gauge symmetry in the Hamiltonian formalism.
Consider a system with a Hamiltonian $H$ which is invariant under the action of some symmetry generators $C_\alpha$:
\begin{equation}
[H,C_\alpha]=0
\end{equation}
Also the space of physical states is restricted to satisfy
\begin{equation}
C_\alpha|\psi\rangle=0
\end{equation}
All observables in such a theory are necessarily also gauge invariant
\begin{equation}
[C_\alpha,O]=0
\end{equation}
Thus an expectation value of any observable in any physical state is given by the formal expression
\begin{equation}\label{element}
\langle O\rangle=\int dx_idy_\alpha \Psi^*(x,y)O\Psi(x,y)
\end{equation}
Here we have intentionally separated the coordinates into two sets $\{x_i\}$ and $\{y_\alpha\}$, since we intend to impose gauge fixing conditions and solve them for $y_\alpha$.  
We want to impose the gauge fixing 
\begin{equation}\label{gaugefix}
G_\alpha(x,y)=0
\end{equation}
Let us {\it a la} Fadeev-Popov multiply the matrix element in eq.(\ref{element}) by unity
\begin{equation}
1=\int d\lambda_\alpha \delta[G_\alpha(x^\lambda,y^\lambda)]|{\rm det}{\delta G_\alpha\over\delta\lambda_\beta}|
\end{equation}
where $x^\lambda$ and $y^\lambda$ as usual are $x$ and $y$ transformed by the gauge transformation with gauge parameters $\lambda$. Just like in the usual Fadeev-Popov approach we can now change variables in the integral $x,y\rightarrow x^\lambda, y^\lambda$. Due to gauge invariance of both the wave function $\Psi$ and the operator $O$, the dependence on $\lambda$ drops out of the integrand, and the $\lambda$ integral gives an irrelevant constant. We thus have
\begin{eqnarray}\label{element1}
\langle O\rangle&=&\int dx_idy_\alpha \Psi^*(x,y)O\Psi(x,y)\delta[G_\alpha(x,y)]{\rm det}{\delta G_\alpha\over\delta\lambda_\beta}|\nonumber\\
&=&\int dx_i \Psi^*(x,y=Y(x))\left[O\Psi(x,y)\right]_{y=Y(x)}|{\rm det}^{-1} {\delta G_\alpha\over\delta y_\beta}{\rm det}{\delta G_\alpha\over\delta\lambda_\beta}|_{y=Y(x)}
\end{eqnarray}
Here $y_\alpha=Y_\alpha(x)$ are solution of the gauge fixing conditions eq.(\ref{gaugefix}). This equation has a simple interpretation. Any matrix element of the type of eq.(\ref{element}) can be calculated as a matrix element on the smaller Hilbert space, the space spanned by coordinates $x_i$ only
\begin{equation}
\int dx_i\mu(x)\bar\Psi^*(x)\bar O(x,p)\bar\Psi(x)
\end{equation}
On this space the norm and the scalar product are defined with a nontrivial measure
\begin{equation}
\mu(x)=|{\rm det}^{-1} {\delta G_\alpha\over\delta y_\beta}{\rm det}{\delta G_\alpha\over\delta\lambda_\beta}|_{y=Y(x)}
\end{equation}
The wave functions and the operators are related to those of the original gauge invariant theory. For the wave functions this correspondence is simple to define
\begin{equation}
\bar\Psi(x)=\Psi(x,y=Y(x))
\end{equation}
For the operators the correspondence is more subtle. It is straightforward if the operator $O$ depends only on coordinates and not on the momenta. In this case, just like for the wave function
\begin{equation}
\bar O(x)=O(x,y=Y(x))
\end{equation}
In general the correspondence is given by
\begin{equation}
\bar O(x,p)\bar\Psi(x)=\left[O\Psi(x,y)\right]_{y=Y(x)}
\end{equation}
The nontrivial statement is that such a correspondence exists for any operator $O$ so that the operator $\bar O$ is the same for all gauge invariant wave functions $\Psi$.

In general finidng this correspondence requires a nontrivial amount of work. We will not delve into the discussion of the most general case, but instead will only consider a simple situation when the constraints $C_\alpha$ as well as gauge conditions $G_\alpha$ are linear functionals of the coordinates and momenta.
In this case the constraints $C_\alpha$ can be solved simply for the momenta conjugate to $y_\alpha$
\begin{equation}\label{co}
p_\alpha=g_\alpha(x_i,y_\beta,p_i)
\end{equation}
We first show how $p_i$ - momenta conjugate to $x_i$,  are represented when they act on $\bar\Psi$. To do this we have to remember that $\Psi$ is only a function of combinations of coordinates $R_i$ which commute with $C_\alpha$. We can then  write
\begin{equation}
{\partial \Psi(x,y)\over \partial x_i}={\partial \Psi\over \partial R_j}{\partial R_j\over\partial x_i}; \ \ \ \ \ \ \ {d \bar\Psi(x)\over d x_i}={\partial \Psi\over \partial R_j}|_{y=Y(x)}{d R_j\over d x_i}={\partial \Psi\over \partial R_j}|_{y=Y(x)}\left[{\partial R_j\over \partial x_i}+{\partial R_j\over \partial y_\alpha}{d Y_\alpha \over d x_i}\right]
\end{equation}
On the other hand the gauge invariance of $R$ means
\begin{equation}
[R_\alpha,p_\beta-g_\beta(x_i,y_\beta,p_i)]=i\left({\partial R_\alpha\over \partial y_\beta}-{\partial R_\alpha\over\partial x_j}{\partial g_\beta\over p_j}\right)=0
\end{equation}
Combining these equations we find
\begin{equation}
{\partial \Psi(x,y)\over \partial x_i}|_{y=Y(x)}=M_{ij}{d \bar\Psi(x)\over d x_j}
\end{equation}
with
\begin{equation}
M^{-1}_{ij}=\delta_{ij}+{\partial Y_\alpha\over \partial x_i}{\partial g_\alpha\over \partial p_j}
\end{equation}
By the generalization of this argument we can show that the same is true for the second derivative of $\Psi$ provided the matrix $M$ is a c-number matrix (which is where the linearity of $C$ and $G$ is important).
\begin{equation}
{\partial^2 \Psi(x,y)\over \partial x_i\partial x_j}|_{y=Y(x)}=M_{ik}M_{jl}{d^2 \bar\Psi(x)\over d x_kdx_l}
\end{equation}
Thus we see that on the reduced Hilbert space the commutators of the coordinates $x_i$ and momenta $p_i$ are modified
\begin{equation}
[x_i,p_j]=-iM_{ij}
\end{equation}
The matrix $M$ is in fact the Dirac bracket  as shown in Appendix B.

The momentum $p_\alpha$ can be eliminated in favor of $p_i$ using the constraint eq.(\ref{co}).
One has to be careful though, as this constrained can only be used when the operator $p_\alpha$ acts on a gauge invariant state. Thus
\begin{eqnarray}\label{howto}
p_\alpha|\Psi\rangle&=&g_\alpha(x,y,p_i)|_{y=Y(x)}|\Psi\rangle;\\
p_\alpha p_\beta|\Psi\rangle&=&p_\alpha g_\beta(x,y,p_i)|\Psi\rangle={1\over 2}\left[\{g_\beta(x,y,p_i),g_\alpha(x,y,p_i)\}-i\left({\partial g_\beta\over\partial y_\alpha}+{\partial g_\alpha\over\partial y_\beta}\right)\right]|_{y=Y(x)}|\Psi\rangle\nonumber\\
p_\alpha p_i|\Psi\rangle&=&p_ig_\alpha(x,y,p_i)|\Psi\rangle=\left[g_\alpha(x,y,p_i)p_i-i{\partial g_\alpha\over\partial x_i}\right]|_{y=Y(x)}|\Psi\rangle\nonumber
\end{eqnarray}

Thus we get the following algorithm to represent an arbitrary operator $O$ on a gauge fixed Hilbert space.. 

1. Order factors of momenta and coordinates such that all the momenta  are to the right,  and all the coordinates to the left.

2. Carefully realize the Gauss law constraints using eq.(\ref{howto}) and its generalizations if necessary.

3. Represent all momenta $p_i$ by derivatives of $\bar\Psi$. In the classical approximation this amounts to replacing the Poisson brackets of $p_i$ by Dirac brackets. In the quantum theory this requires more work, in particular new quantum terms arise which involve derivatives of the Dirac brackets. When constraints are linear, the Dirac bracket procedure is exact also in the quantum case.

4. Finally express all $y_\alpha$ in terms of $x_i$.

This concludes our discussion of gauge fixing in the Hamiltonian formalism. Finally we note that although the Gauss' law in QCD is nonlinear, for the purpose of the present paper the discussion in this appendix suffices. The reason is that our aim was to derive the expression for the $S$-matrix for large fields $b^\pm_i$. In this situation it is sufficient to expand the Gauss' law to first order in fluctuations around the classical fields. In leading order the constraint becomes linear and the procedure described in this appendix applies. Moreover, imposing Dirac brackets is equivalent to operatorially solving the gauge fixing condition, which justifies the quantization procedure used in this paper. Beyond the (semi)classical approximation the qunatization procedure is well defined, but is more subtle as explained above.

\section{Appendix B. Equivalence with Dirac brackets.}
Here we show that the commutation relations discussed in Appendix A are equivalent to Dirac bracket.
Consider the Dirac bracket quantization of the system with constraints 
\begin{equation}
C_\alpha=p_\alpha-g_\alpha(x_i,y_i,p_i)=0; \ \ \ \ \ G_\alpha=y_\alpha-Y_\alpha(x)=0
\end{equation}
The Dirac brackets are defined as
\begin{equation}
\{A,B\}_D=\{A,B\}_P-\{A,c_a\}_PC^{-1}_{ab}\{c_b,B\}_P
\end{equation}
where $c_a$ is the complete set of second class constraints (in our case $C_\alpha$ and $G_\alpha$) and the matrix $C_{ab}$ is defined as
\begin{equation}
C_{ab}=\{c_c,c_b\}_P
\end{equation}
The matrix $C$ is antisymmetric. In our case it has vanishing diagonal blocks, while the off diagonal blocks are
\begin{equation}
L_{\alpha,\beta}=\{C_\alpha,G_\beta\}_P=-\left(\delta_{\alpha\beta}+{\partial g_\alpha\over\partial p_k}{\partial Y_\beta\over\partial x_k}\right)
\end{equation}
Now consider the Dirac brackets between the coordinates and momenta
\begin{equation}
\{p_i,x_j\}_D=-\delta_{ij}-{\partial Y_\alpha\over \partial x_i}L^{-1}_{\alpha\beta}{\partial g_\beta\over\partial p_j}=-\delta_{ij}+{\partial Y_\alpha\over \partial x_i}\left[\delta_{\alpha\beta}-{\partial g_\alpha\over\partial p_k}{\partial Y_\beta\over \partial x_k}+{\partial g_\alpha\over\partial p_k}{\partial Y_\gamma\over \partial x_k}{\partial g_\gamma\over\partial p_l}{\partial Y_\beta\over \partial x_l}-...\right]{\partial g_\beta\over\partial p_j}
\end{equation}
This has to be compared with the result obtained in the text
\begin{equation}
[p_i,x_j]=-iM_{ij}=-i[\delta-{\partial Y_\alpha\over \partial x_i}-{\partial g_\alpha\over\partial p_j}]^{-1}=-i\left[\delta_{ij}-{\partial Y_\alpha\over \partial x_i}{\partial g_\alpha\over\partial p_j}+{\partial Y_\alpha\over \partial x_i}{\partial g_\alpha\over\partial p_k}{\partial Y_\beta\over \partial x_k}{\partial g_\beta\over\partial p_j}-...\right]
\end{equation}
The two expressions obviously coincide. We thus see that our analysis in Appendix A reproduces the Dirac bracket quantization procedure. We note however, that the equivalence with the Dirac bracket quantization only holds on the classical level, or in a fully quantum theory when all the constraints are linear. If the constraints are nonlinear, then the quantum theory involves the Faddeev-Popov determinant and the additional contribution to the measure, which can not be obtained classically. Also if the matrix $M$ depends on fields (which is the case for nonlinear constraints) one has to be careful with the operator ordering and order operators consistently following the technique discussed in Appendix A.


\begin{thebibliography}{99}




\bibitem{JIMWLK} I. Balitsky, {\it Nucl. Phys.}  {\bf B463} 99 (1996); J. Jalilian Marian, A. Kovner and H. Weigert, {\it Phys. Rev.}{\bf D59} 
014015 (1999); 
 A. Kovner, J.G. Milhano and H. Weigert,
{\it Phys.Rev.} {\bf D62} 114005 (2000); 
 Y.~V.~Kovchegov,
  Phys.\ Rev.\ D {\bf 61}, 074018 (2000)
E. Ferreiro, E. Iancu, A. Leonidov, L. McLerran;  
{\it Nucl. Phys.}{\bf A703} (2002) 489.


\bibitem{one} T. Altinoluk, A. Kovner and J. Peresutti, Phys.Lett.B659:144-148,2008.
e-Print: arXiv:0709.0476 [hep-ph]; 


\bibitem{zakopane}
A. Kovner Acta Phys.Polon.B36:3551-3592,2005.
e-Print Archive: hep-ph/0508232

 
\bibitem{klwmij}
A. Kovner and M. Lublinsky; Phys.Rev.D71:085004,2005.
e-Print Archive: hep-ph/0501198


\bibitem{balitsky} I. Balitsky, Phys.Rev.D70:114030,2004.
e-Print: hep-ph/0409314; Phys.Rev.D72:074027,2005.
e-Print: hep-ph/0507237; Nucl.Phys.Proc.Suppl.152:275-278,2006. 


\bibitem{mkw} L. McLerran, A. Kovner and H. Weigert, Phys.Rev.D52:6231-6237,1995.
e-Print: hep-ph/9502289; Phys.Rev.D52:3809-3814,1995.
e-Print: hep-ph/9505320

\bibitem{im} E. Iancu and A. Mueller; Nucl.Phys.A730:460-493,2004.
e-Print: hep-ph/0308315

 
\bibitem{duality} A. Kovner and M. Lublinsky; Phys. Rev. Lett.{\bf 94}, 181603 (2005)

\bibitem{remarks}  A. Kovner and M. Lublinsky; JHEP 0503:001,2005.
e-Print Archive: hep-ph/0502071


\bibitem{foam} A. Kovner, M. Lublinsky and U. Wiedemann, JHEP 0706:075,2007.
e-Print: arXiv:0705.1713 [hep-ph];


\bibitem{long} R.J. Fries, J.I. Kapusta, Y. Li;; e-Print: nucl-th/0604054; D. Kharzeev, K. Tuchin; Nucl.Phys.A753:316-334,2005.
e-Print: hep-ph/0501234; T. Lappi and L. McLerran; Nucl.Phys.A772:200-212,2006.
e-Print: hep-ph/0602189;


\bibitem{glasma} A. Dumitru, F.Gelis , L. McLerran, R. Venugopalan; 
e-Print: arXiv:0804.3858 [hep-ph] 







\end{thebibliography}
\end{document}